\documentclass[12pt]{article}
\usepackage{graphicx}
\usepackage{amsmath,amssymb}
\usepackage{hyperref}
\begin{document}
\title{60 years of attosecond physics at ICPEAC: from collisions to ultrashort pulses}

\author{
Joachim Burgd\"orfer\thanks{burg@concord.itp.tuwien.ac.at}, Christoph Lemell\\
Institute for Theoretical Physics\\
Vienna Univ. of Technology, Vienna, Austria, EU
\and 
Xiao-Min Tong\\
Center for Computational Sciences\\
University of Tsukuba, Tsukuba, Japan}

\maketitle

\begin{abstract}
The field of attosecond physics has seen an almost explosive growth since the early 2000's and represents by now an increasing fraction of contributions to the bi-annual series of International Conferences of Photonic, Electronic, and Atomic Collisions (ICPEAC). The latter is anything but a coincidence as many of the underlying concepts of electronic and photonic dynamics are closely intertwined with atomic-scale collision processes. We illustrate this fruitful connection and its implications with the help of a few prototypical examples of current topical interest.
\end{abstract}

\section{Introduction}

Since the beginning of this millennium the field of ``attosecond physics'' or, more generally, of ``attosecond science'' has seen a near-exponential growth. Taking the number of recent research articles accounted by \textit{Web of Science} with the term ``attosecond'' just in the title as an admittedly superficial indicator, this number rose from near zero before 1995 to 2275 as of July 2019 \cite{WOS}. This disruptive development has its origin in the appearance of novel light sources with pulses as short as $\sim 50$ attoseconds (1 as $=10^{-18}$s) \cite{Krausz09,GBOR}. With the advent of chirped pulse amplification \cite{Mourou85} and the frequency-comb technology \cite{Udem99}, the generation of high-intensity laser beams as well as phase-stabilized few-cycle IR pulses, attosecond pulse trains \cite{Paul01} and isolated attosecond pulses \cite{Drescher01} could be realized and are now routinely employed in many labs around the world.

The importance of well-characterized light fields on the time scale of attoseconds originates from the fact that this is the natural time scale of electronic dynamics in valence shells of atoms, molecules and solids. As already used by Niels Bohr in his atomic model \cite{Bohr13a}, the orbital period of the atomic electron in hydrogen, $T_n=2\pi n^3$ (in atomic units), is $T_1\simeq 150$ as for the ground state (principal quantum number $n=1$). Attosecond pulses thereby offer the opportunity to observe, to interrogate and, eventually, to actively control and manipulate electronic dynamics. A few examples for the realization of this promise involving state-of-the art light fields will be briefly discussed in the following.

\begin{figure}[h]
\centerline{\includegraphics[width=17pc]{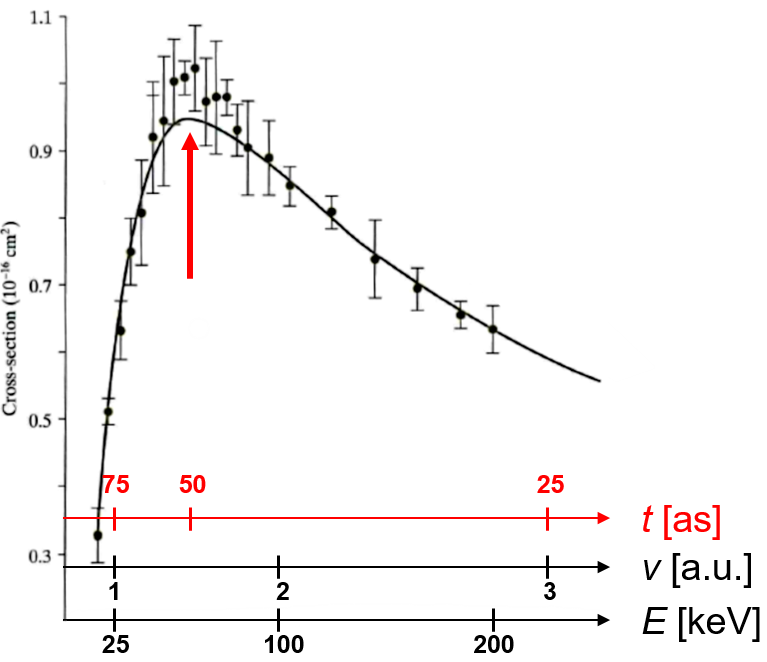}}
\caption{\label{fig1} Excitation cross section for proton impact on hydrogen, H$^+ +$ H(1s) $\to$ H$^+ +$ H($n=2$), as a function of projectile velocity $v$ and collision time $\tau_c=v/2a$ ($a$: atomic radius). Experimental data \cite{Park76}, solid line: atomic close coupling calculation \cite{Ford93}.}

\end{figure}

Exploitation of electromagnetic pulses on the attosecond time scale to induce electronic transitions in atoms, molecules and solids has a long history in the field of atomic collisions for almost a century. Following up on Bohr's work on stopping of charged particles in matter \cite{Bohr13b}, Weizs\"acker \cite{Weizsaecker34} and Williams \cite{Williams35} described in the thirties of the 20th century the interaction of a fast charged particle passing by an atom in terms of a time dependent electromagnetic field pulse of ``virtual photons'' whose duration is on the attosecond scale. Consequently, the probability of Coulomb excitation peaks when the Fourier spectrum of the virtual photon field, $\omega\approx 1/\tau_c=v/2a$ ($v$ is the collision velocity, $a$ the typical linear dimension of the atomic target and $\tau_c$ the collision time), matches the atomic excitation energy $\Delta E = \varepsilon_f-\varepsilon_i$. This celebrated Massey criterion \cite{Massey33} often referred to as the velocity-matching criterion has been one of the corner stones of atomic collision physics, the overarching theme of ICPEAC now celebrating its 60th anniversary. (Sir Harrie Massey was involved in the third edition of ICPEAC in London in 1963). Fig.\ \ref{fig1} illustrates the physics of the virtual photon field and of the resultant Massey criterion using the data for the total excitation cross section of the prototypical collision system of protons colliding with hydrogen (H$^+ +$ H $\to$ H$^+ +$ H($n=2$)). Converting the collision velocity scale to a time scale clearly shows that the excitation probability is highest for a collision time of about 50 as.

Since at that time the rapidly developing accelerator technologies, mostly for nuclear applications, were the driving force for the burgeoning field of atomic collisions, the primary focus was on collision energy or collision velocity rather than on collision time as key parameter for classifying and describing the underlying processes. This may have been one reason why attosecond physics, even though ubiquitously present, was not yet explicitly recognized as such. A deeper reason was likely the conceptual difficulty in mapping time onto a physical variable going back to the early days of quantum physics \cite{Pauli26,Cohen77}. As will be illustrated in the following, recent progress in the field of attosecond physics has helped to explore and to elucidate the role of time in quantum dynamics.

The aim of this presentation given as a plenary talk at ICPEAC XXXI is to demonstrate the close conceptual interconnection between the fields of atomic collisions and of light-field driven attoscience with the help of a few prototypical examples. These examples are chosen to highlight the interplay between photonic and collision processes on ultrashort timescales and are not meant to be representative or a review of the diverse and rapidly growing field of attosecond physics. Atomic units ($e=\hbar=m_e=1$) are used throughout the text unless otherwise stated.

\section{Synthesizing collision-like fields: the half-cycle pulses}

\begin{figure}
\centering{\includegraphics[width=30pc]{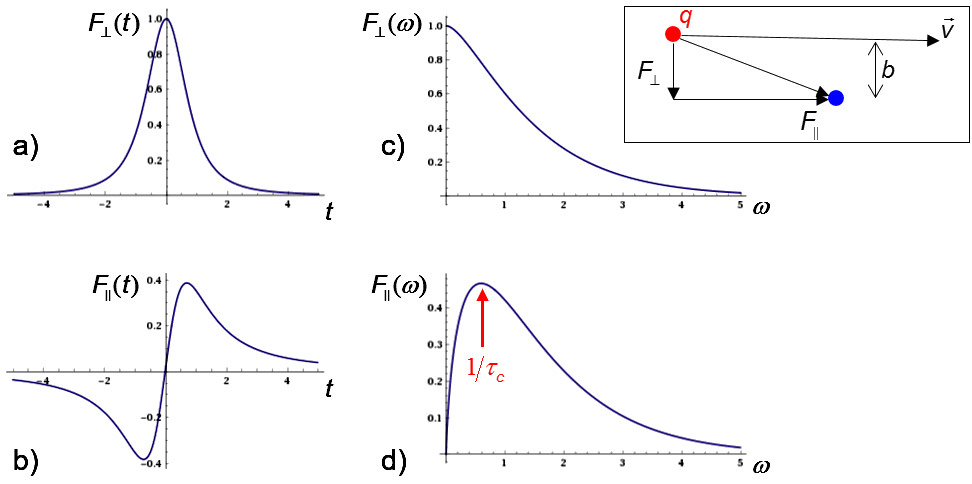}}
\caption{\label{fig2} Temporal evolution of the projectile Couplomb field acting on the target perpendicular (a) and parallel (b) to the collision velocity $\vec v$ (insert: straight-line trajectory with impact parameter $b$). Fourier spectrum of the collisional field pulse perpendicular (c) and parallel (d) to $\vec v$.}
\end{figure}

The temporal shape of the collisional virtual photon pulse (Fig.\ \ref{fig2}) analyzed by Weizs\"acker \cite{Weizsaecker34} and Williams \cite{Williams35},
\begin{eqnarray}
F_\perp(t)&=& q\, \frac{b}{[b^2+(vt)^2]^{3/2}}\, ,\label{WW1}\\
F_\|(t)&=& q\, \frac{vt}{[b^2+(vt)^2]^{3/2}}\, ,\label{WW2}
\end{eqnarray}
($q$: charge of projectile, $b$: impact parameter) form a half-cycle pulse (HCP) perpendicular to the velocity vector (Eq.\ \ref{WW1}, Fig.\ \ref{fig2}a) and a distorted single-cycle pulse along the velocity vector (Eq.\ \ref{WW2}, Fig.\ \ref{fig2}b). For fast collisions ($v\gg 1$ a.u.) the temporal width of the collisional HCP, $\tau_c$, can be as short as a fraction of an attosecond \cite{Moshammer97}. Correspondingly, their Fourier spectra,
\begin{eqnarray}
F_\perp(\omega)&=& \frac{q\omega}{v^2}\cdot K_1(\omega b/v)\, ,\label{WW3}\\
F_\|(\omega)&=& \frac{q\omega}{v^2}\cdot K_0(\omega b/v)\, ,\label{WW4}
\end{eqnarray}
where $K_0$ and $K_1$ are modified Bessel functions of the second kind, extend to the XUV and, for relativistic collisions, well into the X-ray regime. In fact, because of the time (or phase) shift between the two orthogonal components the vectorial field of the passing-by charged particle forms an elliptically polarized half-cycle pulse capable of transferring orbital angular momentum to the electrons of the target. This rotational coupling results in orientation and alignment of the atomic charge cloud and has been a frequently explored topic at many ICPEACs in the 70's and 80's \cite{Fano73,Andersen86}.
Collisions are thus the ideal source of ultrashort pulses in terms of both their absolute duration and the minimum number of cycles subtended. However, in collisions, it is difficult to independently control and vary $v$ and $b$ while suppressing competing processes occurring at close distances, in particular charge transfer.

Synthesizing collision-like pulses under well-controlled conditions, in particular HCPs, was therefore pursued to study the non-linear electronic response to HCPs. Freely propagating HCPs with durations $\tau_\mathrm{HCP}$ of a picosecond or less can be generated using a photoconducting switch triggered by a femtosecond laser \cite{Jones93,You93}. This is still slow compared to the classical orbital period $T_n$ of ground-state atoms ($n=1$) but ultrafast ($\tau_\mathrm{HCP}\ll T_n$) for high Rydberg states. Impulsively driven ultrafast electronic dynamics was therefore first explored in the realm of Rydberg physics \cite{Dunning05}.

One difficulty in using freely propagating HCPs is that they are not inherently truly unidirectional, i.e., do not exactly match Eq.\ \ref{WW1}. They rather comprise a short intense unipolar electric field pulse followed by a weaker pulse of opposite polarity and much longer duration. The resulting pulse shape has been referred to as an ``asymmetric monocycle'' and resembles somewhat Eq.\ \ref{WW2}. The effects of the reverse polarity pulse must be considered in analyzing data. Nevertheless, significant momentum transfer to a Rydberg electron can be achieved because of the very different time scales associated with the unipolar pulse and the ensuing opposite-polarity tail.

Truely unipolar but non-propagating HCPs can be created by applying voltage pulses from a fast pulse generator(s) to an electrode. This approach has the advantage that by using a number of pulse generators in conjunction with pulse combiners and splitters it is possible to engineer complex pulse sequences that can be measured directly using a fast probe and sampling oscilloscope. The minimum pulse durations that can be generated, however, are typically limited to $\gtrsim 0.5-1$ ns meaning that the true impulsive limit can only be reached for atoms with very high $n\gtrsim 300-400$.

In the limit $\tau_\mathrm{HCP}\ll T_n$, a single pulse $\vec F_\mathrm{HCP}(t)$ simply delivers a collision-like impulsive momentum transfer or ``kick''
\begin{equation}
\Delta \vec p=-\int_{-\infty}^\infty F_\mathrm{HCP}(t)\, dt\label{momtrans}
\end{equation}
to the atomic electron. The response of the electron becomes independent of the detailed shape of the pulse as only the integral matters. Classically, the application of such an HCP to an atomic electron with initial momentum $\vec p_i$ and energy $E_i$ results in an energy transfer
\begin{equation}
\Delta E=\varepsilon_f-\varepsilon_i=\frac{(\vec p_i+\Delta \vec p)^2}{2}-\frac{p_i^2}{2}=\frac{\Delta p^2}{2}+\vec p_i\cdot\Delta \vec p\, ,
\end{equation}
where $\varepsilon_f$ is the final electron energy.

\begin{figure}
\centerline{\includegraphics[width=18pc]{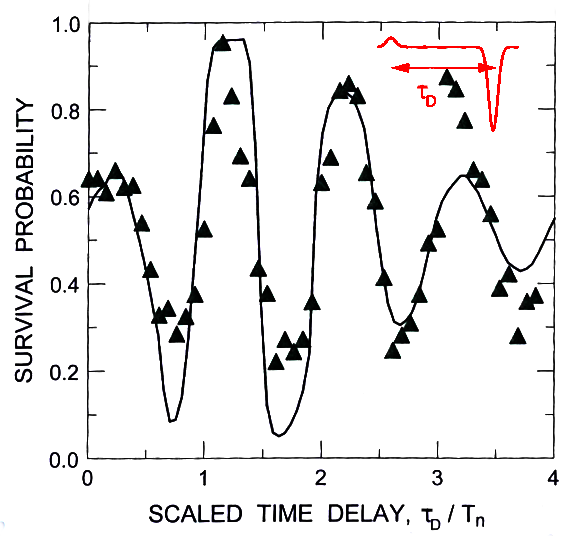}}
\caption{\label{fig3} Measured survival probability for polarized K($n=351$) atoms following application of two oppositely directed HCPs as a function of scaled time delay $\tau_D/T_n$. The HCP sequence is illustrated in the inset. The first HCP delivers a scaled impulse $\Delta p_0=\Delta p/p_n=n\Delta p=-0.1$, the second probe pulse a scaled impulse $\Delta p_0=0.9$. Solid line: results of CTMC simulations (adapted from \cite{Dunning05}).}
\end{figure}

Quantum mechanically, the initial atomic state $|\phi_i\rangle$ is shifted (or ``boosted'') in momentum space by the application of the HCP, $|\psi(t=0)\rangle=|\phi_i^B\rangle=e^{i\Delta\vec p\cdot\vec r}|\phi_i\rangle$. The corresponding expectation values of energy and momentum are
\begin{eqnarray}
\langle \varepsilon\rangle_{t=0}&=&\langle \phi_i^B|H_\mathrm{at}|\phi_i^B\rangle=\langle \phi_i|H_\mathrm{at}|\phi_i\rangle+\frac{\Delta p^2}{2}+\langle \phi_i|\vec p\cdot\Delta\vec p|\phi_i\rangle\, , \\
\langle\vec p\rangle_{t=0}&=&\langle \phi_i^B|\vec p|\phi_i^B\rangle=\Delta\vec p+\langle \phi_i|\vec p|\phi_i\rangle\, ,
\end{eqnarray}
where $H_\mathrm{at}$ is the atomic Hamiltonian. Since for bound states $\langle \phi_i|\vec p|\phi_i\rangle=0$, $\langle \varepsilon\rangle_{t=0}=\varepsilon_i+\Delta p^2/2$, in agreement with the classical prediction. The ``boosted'' final electronic wavefunction can be expanded in terms of the field-free atomic basis as
\begin{equation}
|\psi(t)\rangle = \sum_j a_j(0) e^{-i\varepsilon_j t}|\phi_j(0)\rangle \label{Ryd}
\end{equation}
and comprises a coherent superposition of many states having a broad distribution in $j=(n,\ell,m)$. Eq.\ \ref{Ryd} represents a nonstationary wavepacket whose evolution is governed by the atomic, i.e., field-free Hamiltonian $H_\mathrm{at}$. The evolution of $|\psi(t)\rangle$ has been observed by applying a second probe HCP of opposite polarity after some time delay $\tau_D$ (inset Fig.\ \ref{fig3}). This pulse sequence resembles, in fact, the longitudinal component of the collisional pulse (Eq.\ \ref{WW2}). Unlike the collisional pulse, however, it features exquisite tunability of amplitude and delay between the two HCPs. Pronounced quantum beats are observed in the survival probability of K($n=351$) Rydberg atoms \cite{Arbo03} well reproduced by quasi-classical CTMC simulations (Fig.\ \ref{fig3}). This close classical-quantum correspondence for beats is an instructive illustration of Bohr's correspondence principle and points to periodic charge fluctuations of the revolving electron as their origin.

HCP-driven Rydberg atoms, discussed at several ICPEACs (see, e.g., \cite{Yoshida04}), can be viewed as a precursor to current attosecond pulse interactions with atoms. Several fundamental concepts related to the non-linear electronic response to ultrashort pulses could be explored, however, only on the nano-to-picosecond time scales and not yet for valence electrons. Only with the advent of attosecond pulses, watching and probing electronic motion near the electronic ground state became feasible.

\section{Attosecond time delays}\label{sect3}

The availability of carrier-envelope phase (CEP) controlled IR pulses and of attosecond pulses has allowed to ``clock'' fundamental electronic processes. Observing quantum dynamics in the time domain, often referred to as attosecond chronoscopy \cite{Pazourek15}, can be viewed as source of information complementary to that of spectroscopy.

One of the first major breakthroughs of attosecond chronoscopy was the observation of a finite time delay of photoemission relative to the arrival of the crest of the ionizing XUV pulse \cite{Schultze10}. These time delays, typically of the order of about ten attoseconds, are directly related to the Eisenbud-Wigner-Smith (EWS) time delay \cite{Eisenbud48,Wigner55,Smith60} originally introduced for resonant elastic scattering,
\begin{equation}
\tau_\mathrm{EWS}=i\hbar S^\dagger(E)\frac{\partial}{\partial E}S(E)\, ,\label{EWS}
\end{equation}
in terms of the scattering matrix $S(E)$. Felix Smith, member of the organizing committee of the first ICPEAC in 1958, presented the time-delay operator in matrix form for multi-channel problems at ICPEAC II in 1961 \cite{Smith61}. The Hermitian operator $\tau_\mathrm{EWS}$ features real eigenvalues, the eigentime delays, providing an explicit and constructive example of a bona-fide physical observable of quantum theory. The EWS time delay (Eq.\ \ref{EWS}) satisfies the correspondence principle of quantum to classical mechanics. $\tau_\mathrm{EWS}$ can be shown to converge in the limit of small deBroglie wavelength $\lambda_\mathrm{dB}\to 0$ to the classical time delay $\tau_\mathrm{cl}$ given, e.g., for a 1d scattering problem for a potential $V(x)$, by
\begin{equation}
\tau_\mathrm{cl}=\lim_{X\to\infty}\left[\int_{-X}^X dx\, \frac{1}{\sqrt{2m(E-V(x))}}-2\,\frac{X}{\sqrt{2mE}}\right]\, .
\end{equation}

Measurement of EWS time delays for photoionization has been realized by two alternative pump-probe settings involving an attosecond pulse (or pulse train) as a pump ionizing the atom and a few-cycle IR pulse as probe, both temporally near-perfectly correlated with each other through the process of high-harmonic generation (HHG). An intense IR pulse generates high-frequency radiation by strongly non-linear up-conversion \cite{McPherson87,Ferray88,Huillier93,Lewenstein94}. The underlying mechanism is nicely captured by the instructive semiclassical three-step model \cite{Kulander93,Corkum93} of tunneling ionization, acceleration by the strong IR field, and rescattering at the ionic core resulting in coherent radiative recombination. It reveals the close connection to the physics of electron-ion recombination, a major topic at many editions of ICPEAC's (see, e.g., \cite{Walton70,Muller91}).

The same IR pulse, suitably attenuated, provides in addition a temporally well correlated moderately strong IR probe field ($\sim 10^{11}$ W/cm$^2$ for streaking). The latter ponderomotively shifts the momentum of the outgoing photoelectron $\vec p_f$ relative to its field-free value, $\vec p_0$, by the vector potential of the IR field $\vec A_\mathrm{IR}(t)$,
\begin{equation}
\vec p_f(t)=\vec p_0-\vec A_\mathrm{IR}(t)/c\label{eq12}
\end{equation}
at the instant of time $t$ of its appearance in the continuum. By varying the temporal overlap (or relative phase) between the IR and the attosecond XUV pulses and, hence, by the sinusoidal modulation of the energy of the outgoing electron $E(t)=p_f^2(t)/2$, this attosecond streaking protocol \cite{Schultze10,Itatani02} allows to time the photoemission with attosecond precision. Attosecond streaking can be viewed as a ``classical clock'' mapping time onto energy. It can also be viewed as the (multi-photon) classical limit of a ``quantum clock'', the interferometric RABBITT technique (reconstruction of attosecond bursts by interference of two-photon transitions \cite{Paul01,Veniard96,Kluender11}). The latter utilizes one-photon transitions between continuum states induced by weaker IR pulses to determine phase differences in side bands between the photoelectron wavepackets ejected by adjacent odd harmonics of the XUV attosecond pulse train.

It is worth recalling that the current attosecond streaking technique closely mirrors a streaking method on the femtosecond scale employed in the field of atomic collisions then referred to as post-collision interaction (PCI) effect. Numerous PCI studies were presented at ICPEACs in the 80's. Streaking by PCI was, for example, used to time resolve the collective motion of autoionizing states coherently excited in Li$^+ +$ He $\to$ He$^{**} +$ Li$^+$ collisions \cite{Straten86,Burgdoerfer88}. The energy of the electron emitted by the autoionizing states (lifetime $\sim 20$ fs) was modulated relative to its field-free value by its Coulomb interaction $\sim 1/R(t)$ with the slowly receding Li$^+$ projectile still in the vicinity of the target when autoionization occurs. Thereby, the emission time of the electron was mapped onto its energy shift relative to the kinetic energy of autoionization electron in free space.

\begin{figure}
\centerline{\includegraphics[width=20pc]{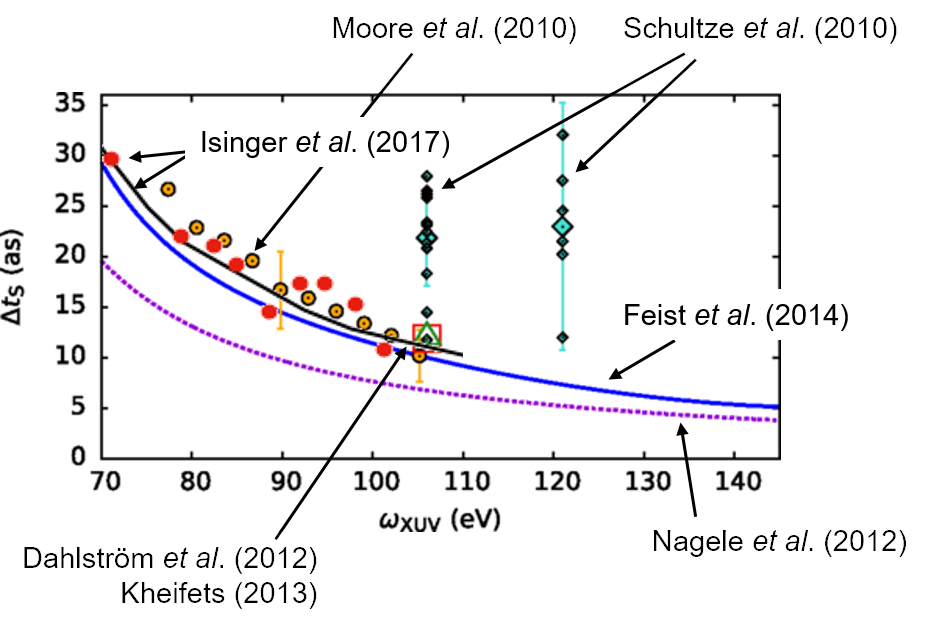}}
\caption{\label{fig4} Relative time delay between the photoelectron emitted from Ne(2s) and from Ne(2p). Experiments \cite{Schultze10,Isinger17} and theory \cite{Schultze10,Moore11,Dahlstroem12,Kheifets13,Feist14,Isinger17}}
\end{figure}

The pioneering attosecond streaking experiment \cite{Schultze10} for the relative EWS time delay between photoionization of the 2s and 2p electrons of neon by photons with energies of about 110 eV yielded a time delay of $\Delta \tau_\mathrm{2s-2p}\approx 20$ as. Accompanying calculations \cite{Schultze10} as well as a large number of subsequent theoretical studies \cite{Moore11,Dahlstroem12,Kheifets13,Feist14} could confirm that the formation of the outgoing 2p wavepacket is, indeed, delayed relative to the 2s wavepacket but found considerably smaller values (Fig.\ \ref{fig4}). As attosecond streaking features excellent time resolution but only limited energy resolution given by the width of the streaking modulation (see Eq.\ \ref{eq12}), $\delta E= |p_0 A_\mathrm{IR}|$, the streaking trace of the main line of the 2s electron overlaps with the emission from shake-up transitions accompanying the 2p emission which may distort the observed time delay \cite{Feist14}. Unlike for neon, photoemission from helium allows resolving shake-up transitions from main lines by streaking. Measurements and calculations demonstrated with one attosecond precision large differences between $\tau_\mathrm{EWS}$ of the main line and shake-up transitions \cite{Ossiander17}.
Exploiting the considerably higher energy resolution of RABBITT, Isinger \textit{et al.}\ \cite{Isinger17} could subsequently determine the relative time delay between the 2s and 2p main lines of neon without distortion by shake-up admixture and found good agreement with theory \cite{Moore11,Dahlstroem12,Kheifets13,Feist14} at somewhat lower photon energies.

Extracting $\tau_\mathrm{EWS}$ for photoionization from attosecond streaking or RABBITT measurements requires the correction by an additional time shift, $\tau_\mathrm{CLC}$, due to the Coulomb-laser coupling (CLC) for streaking or $\tau_\mathrm{CC}$ due to IR induced one-photon continuum-continuum (CC) transitions for RABBITT. Both $\tau_\mathrm{CLC}$ and $\tau_\mathrm{CC}$ are of the same origin, the interaction of the IR probe field with the outgoing electron in the atomic Coulomb field and their values have been found to be nearly identical, $\tau_\mathrm{CLC}\simeq\tau_\mathrm{CC}$ \cite{Pazourek15,Nagele11,Dahlstroem13}. Only recently, these additional time shifts often referred to as measurement related effects have been recognized to be of the same physical origin as $\tau_\mathrm{EWS}$ for photoionization, i.e., for a bound-continuum transition. $\tau_\mathrm{CC}$ (or $\tau_\mathrm{CLC}$)  can be viewed as the EWS time delay for continuum-continuum scattering by absorption or emission of IR photons \cite{Fuchs19}.

In the current generation of experiments, only relative time delays either between different ionization processes of the same atomic species (e.g., Ne(2p) vs.\ Ne(2p) ionization) or different atoms in a gas mixture (e.g., pairs of rare gases \cite{Palatchi14}) are accessible. To establish an absolute time scale for delays, the reference to (virtually) exact solutions of the Schr\"odinger equation is required that has become available, e.g., for helium \cite{Pazourek12,Pazourek13}.

Timing of the photoelectric effect, i.e., the determination of ``time zero'' for emission from solid surfaces poses an even bigger challenge as it represents a true many-body problem in an extended system. Observation of the photoelectron implies tracing out the ``environmental'' degrees of freedom of such a many-body system as concomitant electronic excitation or phonon scattering are not observed. The resulting time delay is therefore associated with partially decoherent dynamics rather than a fully coherent wavepacket. The multi-step process of primary excitation to the conduction band, subsequent transport and scattering and, eventually, transmission through the solid-vacuum boundary is accessible to classical transport simulations \cite{Lemell09}. Time resolved photoemission depends not only on the electronic structure of the topmost layers but, at the same time, also probes the non-linear electronic response of the target surface to (moderately) strong IR fields and the electronic transport on the attosecond time scale.

\begin{figure}
\centering{\includegraphics[width=30pc]{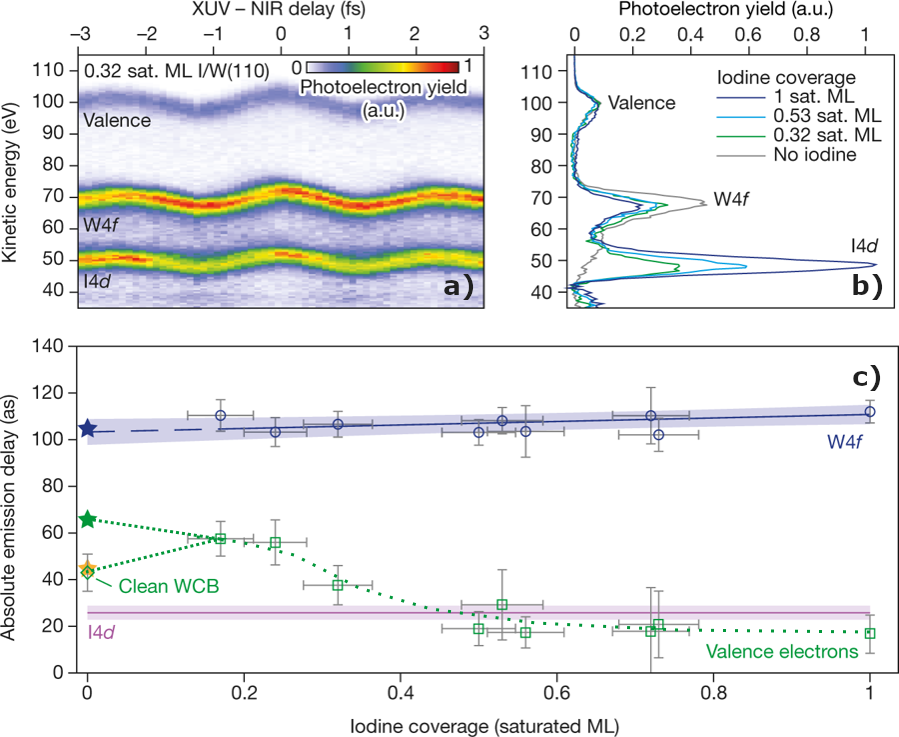}}
\caption{\label{fig5} Measurement of absolute time delay of photoemission from tungsten using iodine as chronoscope. a) streaking trace b) photoelectron spectrum c) delay of the core level W4f and valence band as a function of the coverage of the W surface by chronoscope atoms. Delay of the W conduction band (WCB) for a clean surface is marked. Stars result from transport simulations: blue (W4f), green (WCB without surface states) and yellow (WCB including surface states). Dotted lines to guide the eye. (adapted from \cite{Ossiander18}).}
\end{figure}

The first measurement of the relative timing between emission from the conduction band (WCB) and from the 4f level (W4f) of tungsten \cite{Cavalieri07} resulted in remarkably long time delays of $\sim 100$ as and stimulated a large number of theoretical investigations (e.g.\ \cite{Lemell09,Kazansky09,Zhang09}). The origin of such time delays has remained an open question for a long time. Considerable progress has been made by controlled deposition of Mg adlayers on the W surface allowing for the accurate relative timing between the 2p-core level of Mg adlayers and the substrate 4f level of W \cite{Neppl15}. A large fraction of the relative time delay could be unambiguously attributed to the transport from the substrate through the top adlayer. Very recently, the absolute timing of the photoelectric effect could be experimentally determined for the first time \cite{Ossiander18}. To this end, iodine was used as an atomic chronoscope. By a reference measurement of the relative time delay between the chronoscope atom and helium for which the absolute time delay $\tau_\mathrm{EWS}$  can be determined ab-initio with (sub)attosecond precision, the time delay for 4d emission of the chronoscope iodine atom can be placed on an absolute scale. In turn, by depositing I on the tungsten surface, the absolute timing of both 4f core level and conduction band emission could be determined (Fig.\ \ref{fig5}).
Furthermore, quenching the contribution from localized surface states by increasing the surface coverage by adsorbate atoms results in a non-monotonic time delay for conduction band electrons. This observation allowed to identify the important contribution of surface states which feature a very short time delay $\tau_\mathrm{SS}\sim 10$ as. Surface states are, in part, responsible for the short mean time delay of the conduction band of clean W of $\tau_\mathrm{CB}\sim 40$ as. These results highlight the sensitivity of the observable time delay to both the local electronic structure and to transport.

\section{Quantum beats and decoherent dynamics}

Another class of time resolved observables is associated with the appearance of quantum beats. A sudden perturbation of the initial state creates a wavepacket composed of a coherent superposition of excited eigenstates (see Eq.\ \ref{Ryd}). An ultrashort pulse, in particular an attosecond pulse with duration $\tau_p$ of a few hundred attoseconds causes such impulsive perturbation thereby coherently exciting many electronically excited bound states and continuum states within the reach of its spectral width $\sim\hbar/\tau_p$ of several eV. The reduced density matrix of the system resulting from tracing out unobserved degrees of freedom (the ``environment'' $E$) \cite{Blum81},
\begin{equation}
\rho(t) = \mathrm{Tr}_E\left(|\psi(t)\rangle\langle\psi(t)|\right)\, ,
\end{equation}
describes the time evolution of the resulting partially decoherent wavepacket \cite{Gabriel69,Bosse74}. Attosecond/IR pump-probe pulse sequences are well suited to generate and to observe the ensuing (de)coherent electronic dynamics that manifests itself in (damped) oscillations, the quantum beats, in, e.g., the one-particle density $\langle\vec r|\rho(t)|\vec r\rangle$ (see also section 2).

Quantum beat spectroscopy has a long history in the field of electronic and atomic collisions going back (at least) to beam-foil spectroscopy in the late 60's \cite{Bashkin68}. Fast atomic or ionic species were impulsively and coherently excited near the exit surface after traversing self-supporting thin carbon foils. Quantum beats were monitored in the time resolved down-stream photoemission from fast projectiles with the time-resolution severely limited by the spatial resolution of the detection. Accordingly, beats on the nanosecond (or GHz) scale such as fine-structure quantum beats in light ions \cite{Andrae74} and Lamb shift s-p coherence beats in hydrogenic systems \cite{Sellin79,Burgdoerfer79} could be resolved. Collisions are an ideal tool for providing impulsive and sudden perturbations and, thus, for generating coherences over a remarkably wide range of energy and time scales extending from nanoseconds for fine-structure quantum beats in hydrogen to the (sub-) attosecond time scale for quasi-molecular X-ray emission in energetic highly-charged ion-atom collisions \cite{Schuch88}.

The availability of well-phase controlled XUV/IR pump-probe pulse sequences on the (sub-) femtosecond time scale has opened novel opportunities to generate and probe quantum beats in electronic systems on ultrashort time scales. One pioneering example was the observation of fine structure quantum beats in Kr$^+$ by attosecond transient absorption (ATA, \cite{Goulielmakis10}) where the p$_{1/2}$-p$_{3/2}$ beat period is 6.2 fs rather than $\sim 0.12$ ns in hydrogen. Another is the charge migration in molecules \cite{Cederbaum99,Calegari14,Kraus15}, i.e., oscillations in the electron or hole density spreading over the molecule due to the impulsive excitation of a coherent superposition of many-electron states. These oscillations in the electronic system are, in general, fully reversible. A net irreversible charge transport occurs only through a ``collapse of the wavepacket'', a decoherent process induced, e.g., by coupling to vibronic degrees of freedom resulting in molecular rearrangement or dissociation.

\begin{figure}
\centering{\includegraphics[width=30pc]{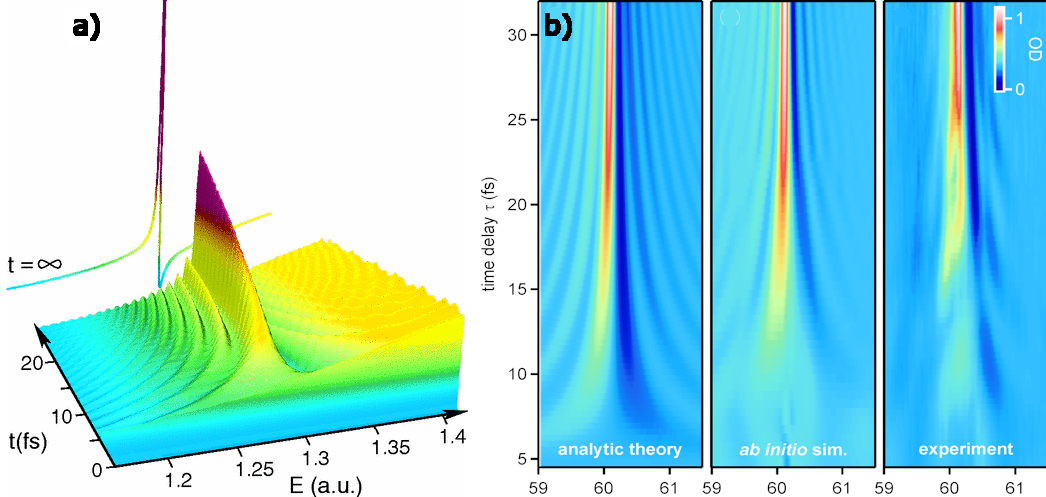}}
\caption{\label{fig6} Build-up of Fano resonance in the time domain. a) Evolution of the photoelectron spectrum and asymptotic Fano profile, b) comparison between experiment, ab-initio simulation and simplified analytical model (adapted from \cite{Argenti13,Kaldun16}).}
\end{figure}
The build-up of a Fano resonance in the time domain gives rise to quantum beats in the continuum \cite{Wickenhauser05,Argenti13} by interference between the direct photoionization path with an amplitude $\sim \langle c|\vec r|i\rangle$ and the indirect ionization path through the quasi-bound resonance ($|R\rangle$) $\sim \langle R|\vec r|i\rangle$ which subsequently decays to the continuum state $|c\rangle$ via autoionization $\sim \langle c|V_{ee}|R\rangle$. The beat amplitude at energy $E$ and time $t$ after impulsive excitation is of the form
\begin{equation}
A(E,t)\sim e^{-\Gamma/2\, t}\cos[(E-E_R)t]\, .
\end{equation}
The crests of the beat ``wave'' forms hyperbolas with $(E-E_R)t=n\pi$ (Fig.\ \ref{fig6}) which converge to the energetic position of the resonance $E_R$ in the limit $t\to\infty$ while the amplitude is damped by the finite lifetime of the resonance ($\sim 1/\Gamma$). Quantum beats in autoionization provide a prototypical example of intrinsic partially decoherent dynamics. The recent first experimental observation of the build-up of Fano resonances in helium by ATA \cite{Kaldun16} and by RABBITT \cite{Gruson16} has been a significant achievement of attosecond physics. It provided novel insights into the temporal evolution of wavepackets in the structured continuum and direct observation of quantum beats otherwise not visible in the well-established stationary ($t\to\infty$) Fano profile of the resonance.

\begin{figure}[h]
\centerline{\includegraphics[width=20pc]{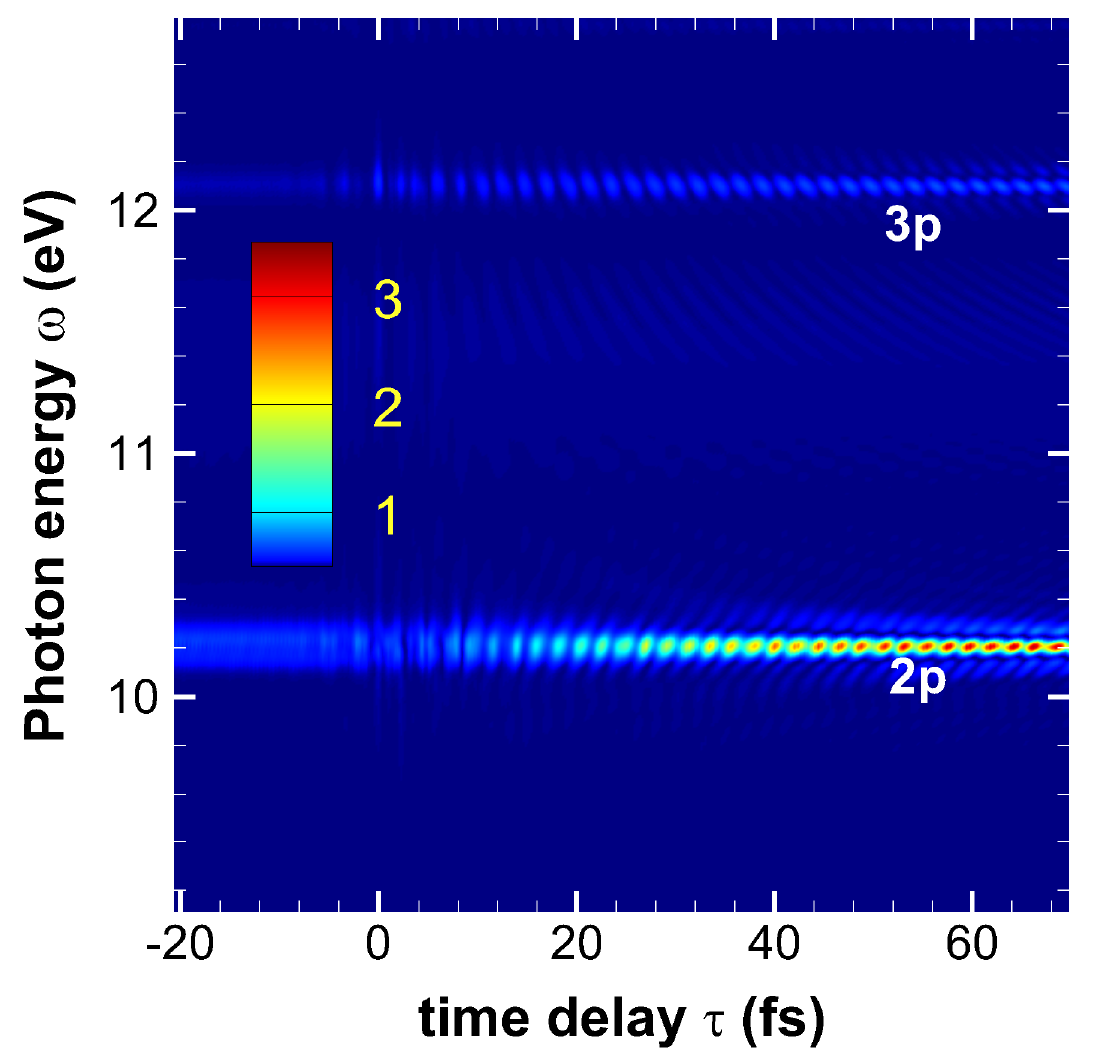}}
\caption{\label{fig7} VUV emission spec\-trum as function of time delay between the OAS pump ($I=5\times 10^{13}$ W/cm$^2$) and OAS probe pulse ($I=2\times 10^{12}$ W/cm$^2$). The Ly$_\alpha$ and Ly$_\beta$ emission lines are labelled by the initial state of the transition (from \cite{Jiang19}).}
\end{figure}

The recent break-through in synthesizing intense optical attosecond (OAS) pulses \cite{Hassan16} resembling half-cycle pulses with an effective temporal width below one femtosecond and a spectral distribution extended from $\sim 1$ eV to about $\sim 4$ eV has enabled for the first time to impulsively generate electronic bound-state wavepackets in atoms involving electronic inter-$n$ coherences without being overshadowed by ionization. With a time-delayed weaker optical attosecond pulse as a probe, the electronic wavepacket motion can be mapped onto a quantum beat signal in photoemission from the excited states \cite{Jiang19}. First experimental tests are underway \cite{Goulielmakis19}. The Ly$_\alpha$ and Ly$_\beta$ emission spectra of hydrogen simulated for an OAS pump-probe pulse sequence as a function of delay time (Fig.\ \ref{fig7}) displays, indeed, an intricate beat pattern that results from strong-field induced interferences between a multitude of excitation paths.

\section{Time resolved holography and diffraction}

Tunneling ionization by a strong IR pulse creates a time-periodic sequence of ionization bursts confined to a small fraction of the optical cycle each emitted near a maximum of the driving field. This sequence of bursts gives rise to an ionization spectrum with approximately equispaced peaks in energy \cite{Kruit83} with the peak spacing given by the IR photon energy often referred to as above threshold ionization (ATI). It was realized early on \cite{Kuchiev87} that ATI can be described by a semiclassical two (or three) step model that could also account for HHG (see Sect.\ \ref{sect3}), however, with elastic electron-ion scattering rather than radiative recombination as the third step. While the ATI peaks are, semiclassically, the signature of temporal intercycle interferences, additional modulations not equispaced in energy result from intracycle interferences \cite{Arbo10}.

\begin{figure}
\centering{\includegraphics[width=30pc]{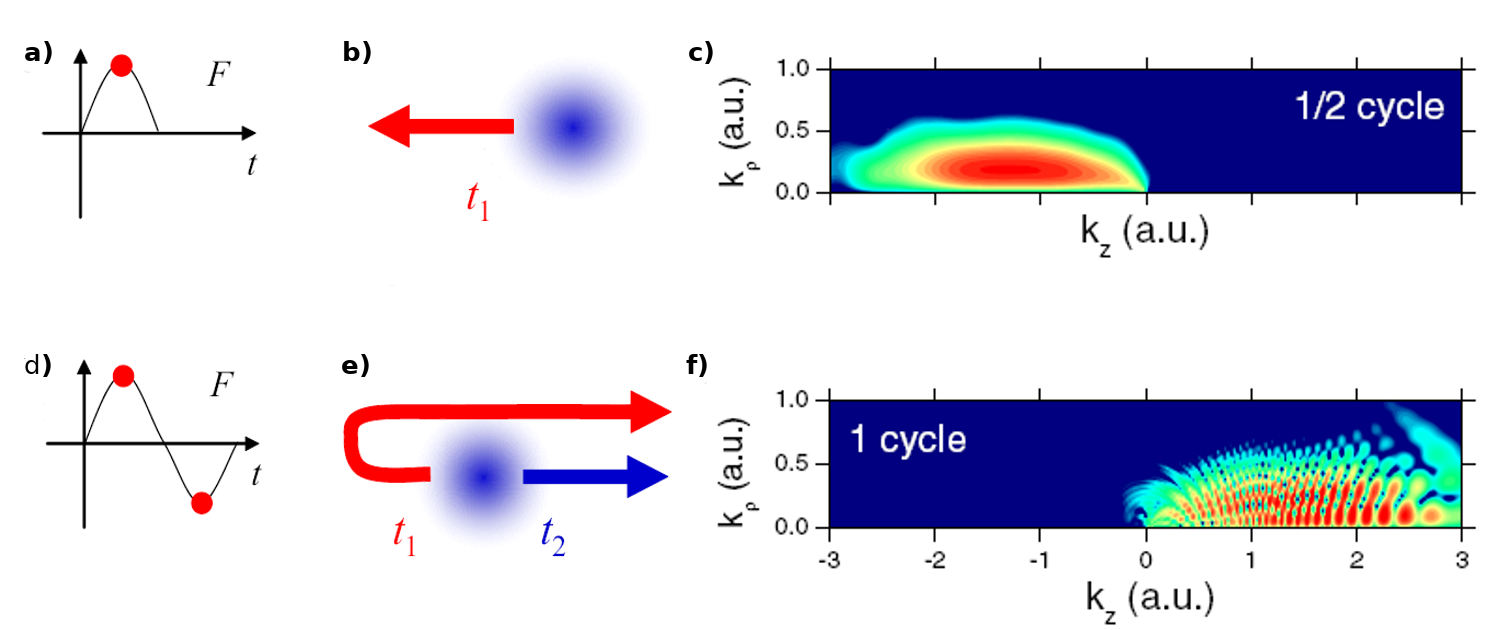}}
\caption{\label{fig8} Temporal double slit for a single-cycle IR pulse ionizing hydrogen ($I=2\times 10^{14}$ W/cm$^2$). a-c: first half cycle. a) temporal distribution of the electric field $F(t)$ with maximum of tunneling ionization at $t=t_1$. b) orientation of unidirectional emission c) resulting vectorial momentum distribution, $P(k_z,k_\rho)$, along ($k_z$) and perpendicular ($k_\rho$) to the laser polarization axis. d-f: as a-c but after the full cycle featuring the rescattered first and a tunnel-ionized second ionization burst. The resulting diffraction pattern can be viewed as a hologram of the time-dependent potential landscape traversed by the wavepacket liberated during the first half cycle between times $t_1$ and $t_2$ (adapted from \cite{Arbo06}).}
\end{figure}
A well-controlled few-cycle IR pulse generates only few ionization bursts each of which extending typically only $\sim 100$ as. Their relative strength can be controlled by the carrier-envelope phase (CEP). For a pulse with effectively only two subsequent ionization bursts, interferences in the photoelectron spectrum signify a temporal double slit \cite{Lindner05} with spacing between the slits of $T/2$ and a width of each slit of $\sim 100$ as. The resulting 2d momentum distribution of the outgoing electron along the laser polarization axis, $p_\|$, and perpendicular to it, $p_\perp$, displays a remarkably complex diffraction pattern (Fig.\ \ref{fig8}f) that carries both \AA ngstrom-scale spatial and attosecond-scale temporal information. The interference of two electron microbursts emitted from an atomic point source can be viewed as holographic image of the potential landscape the receding electron traverses \cite{Arbo06}. This can be easily visualized by considering a single optical cycle consisting of two HCPs of opposite polarity (Fig.\ \ref{fig8}a,d) resembling the collisional field (c.f.\ Fig.\ \ref{fig2}). After the first half-cycle the tunneling ionized 2d momentum distribution is smooth and structureless centered about the mean momentum $\langle p_\|\rangle$ corresponding to the momentum transfer by a ``kick'' (see Eq.\ \ref{momtrans} and Fig.\ \ref{fig8}c). The second HCP reverses the direction of this wavepacket and, moreover, creates a second ionization burst traveling in the same direction (Fig.\ \ref{fig8}e). This second wavepacket serves as the reference wave for the hologram (Fig.\ \ref{fig8}f). The information of the potential landscape, in the present case of the atomic hydrogen Coulomb field and the IR field, traversed by the first wavepacket between the two subsequent field extrema is imprinted on the phase of the rescattered wavepacket. The development of the full potential of this holographic imaging technique with sub-fs temporal and \AA ngstrom-scale spatial resolution, e.g., for molecules, is still in its infancy. First proof-of-principle experiments for metastable xenon strong-field ionized by a mid-IR free electron laser (FEL) pulse \cite{Huismans11} and for simple molecules \cite{Wolter16} have demonstrated its feasibility.

The maximum energy of the redirected and returning electron is of the order of $3U_p$ where $U_p=F_0^2/(4\omega_\mathrm{IR}^2)$ ($F_0$ peak field strength) is the ponderomotive energy associated with the quiver motion of a free electron in the laser field. For large $U_p$, i.e., strong mid-IR pulses, the returning wavepacket can be approximated by a plane wave of a free electron. Accordingly, the interaction with the residual ion can be viewed as elastic or inelastic electron-ion scattering and, for HHG, as radiative recombination. This is the conceptual underpinning of the three-step model for HHG \cite{Huillier93,Lewenstein94,Kulander93,Corkum93} and of the quantitative rescattering theory \cite{Morishita08,Lee09}. It also reveals another close link to the field of electronic and atomic collisions. Cross sections for electron diffraction, electron-impact ionization and radiative recombination can be directly employed in estimates for strong-field processes. As the electron ``beam'' in the rescattering process is temporarily confined to sub-femtoseconds, ultrafast diffraction and microscopy are coming into reach.

\section{Outlook}

With further progress in the development of novel light sources, the rapidly developing field of attosecond physics is expected to expand and widen its scope. It is tempting to speculate which directions this field will take in the future. Certainly, Yogi Berah's famous dictum ``\textit{It is always difficult to make predictions -- in particular about the future}'' applies here. While some of the destinations are already well within reach, others may be longshots.

Attosecond chronoscopy is increasingly complementing conventional spectroscopy, providing information not easily accessible in the frequency domain. One prominent example is relaxation and decoherence dynamics \cite{Pabst11}. Analyzing the ``arrow of time'' of irreversibility is greatly facilitated by observing processes in the time domain and likely to become an important tool, in particular for more complex systems \cite{Jordan15}.

Another topic already well underway is ultrafast magnetic and spin dynamics as well as dichroism \cite{Beaulieu17} accessible now with cirularly polarized attosecond XUV pulses. First observations of ultrafast demagnetization on the (sub-) femtosecond time scale have been reported \cite{Siegrist19,Dewhurst18}. The formation of structured attosecond light beams carrying high orbital angular momenta \cite{Popminchev15} and electron vortices by broad-band circular attosecond pulses \cite{Ngoko15} appear to be in reach.

One obvious direction for future progress will be the further increase in intensity of attosecond XUV pulses so that highly non-linear attosecond processes such as multi-photon ionization and attosecond pump--attosecond probe scenarios can be realized. Further improvement of HHG sources as well as sub-fs pulse control of FEL will be key. One challenging long-term goal would be to reach the high-intensity high-frequency regime of atomic stabilization where the intense light field inhibits rather than stimulates photoelectron emission \cite{Vos92,Jiang18}.

Another tempting direction to be pursued is the quest for even shorter pulses on the sub-attosecond time scale reaching $\sim 10-100$ zeptoseconds (1 zs $=10^{-21}$ s). While this has already been accomplished by high-energy collisions \cite{Moshammer97}, HHG by mid-IR pulses reaching keV photon energies \cite{Popminchev12} provide the first steps towards its realization for light fields. It is, however, still a wide open question which novel physical observable can be explored with zeptosecond pulses.\\[.5cm]
{\bf Acknowledgments:}\\
JB and CL were supported by Austrian Science Fund FWF (Proj. Nos. SFB-F41, W1243), XMT was supported by a Grants-in-Aid for Scientific Research (JP16K05495) from the Japan Society for the Promotion of Science.

\end{document}